\documentclass[a4paper]{jpconf}
\usepackage{graphicx}
\usepackage{hyperref}
\usepackage{amsmath,amssymb}
\newcommand{\be}{\begin{equation}}
\newcommand{\ee}{\end{equation}}
\newcommand{\bea}{\begin{eqnarray}}
\newcommand{\eea}{\end{eqnarray}}

\def\3i{\int\!\!\!\int\!\!\!\int}
\def\2i{\int\!\!\!\int}

\begin{document}
\title{Graviton self-energy from worldlines}

\author{Fiorenzo Bastianelli and Roberto Bonezzi}

\address{Dipartimento di Fisica ed Astronomia, Universit{\`a} di Bologna and\\
INFN, Sezione di Bologna, via Irnerio 46, I-40126 Bologna, Italy}

\begin{abstract}
Worldline approaches, when available, often simplify and make more efficient the calculation of various observables
in quantum field theories. In this contribution we first review the calculation of the graviton self-energy
due to a loop of virtual particles of  spin 0, 1/2 and 1, all of which have a well-known worldline description.
For the case of the graviton itself, an elegant  worldline description is still missing, though one can still  describe it
by constructing a  worldline representation of the differential operators that arise in  the quadratic approximation
of the Einstein-Hilbert action. We have recently analyzed the latter approach, and we use it here to calculate
the one-loop graviton self energy due to the graviton itself in this formalism.

\end{abstract}

\section{Introduction}
The worldline path integral formulation of quantum field theories provides an alternative efficient method for computing Feynman diagrams,
especially in the one-loop approximation, see ref. \cite{Schubert:2001he} for a review.
The worldline method was of particular interest to Victor, who has given various contributions to the subject. In particular, in \cite{Bastianelli:2007jv}
he contributed in analyzing the  consequences of the one-loop graviton-photon mixing in a constant electromagnetic
background \cite{Bastianelli:2004zp}. The Feynman diagram describing the process is drawn in figure 1,
\begin{figure}[h]
\begin{center}
\includegraphics{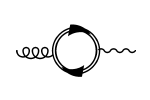}
\end{center}
\caption{\label{fig-uno}One-loop correction to the graviton-photon  mixing due to a virtual loop of an electron in a constant magnetic field.}
\end{figure}
and constitute a prime example where the worldline approach has been used with great efficiency.
The final result of that analysis was that, even for the strong magnetic fields present in our universe,
 the one-loop contribution amounts to  no more than a few percent of the tree level mixing. However, 
although it is numerically only a small correction to the tree-level
amplitude, unlike the tree-level amplitude the one-loop conversion of
photons into gravitons in a magnetic field leads to dichroism, and
surprisingly for the relevant range of parameters is even the largest
standard model contribution to dichroism, as shown in \cite{Ahlers:2008jt}.

Akin to the diagram of the graviton-photon mixing, there are the graviton self-energy diagrams
that take place in flat spacetime  (without the need of additional background fields).
Those that are due to virtual particles of spin 0, 1/2 and 1 may be depicted as in figure 2,
\begin{figure}[h]
\begin{center}
\includegraphics{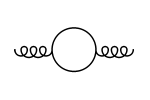}
\end{center}
\caption{\label{fig-due}One-loop correction to the graviton self-energy due to particles of spin 0, 1/2 and 1 circulating in the loop.}
\end{figure}
and can be computed  successfully with worldline methods.
Indeed, a worldline description of particles of spin $s\leq 1$ coupled to external gravity is easily achievable, as we are going
to review in the following. In particular, we outline the essential steps of the worldline calculation of the self-energy diagrams.

The description of the graviton itself is more problematic. One would like to compute in the worldline approach the diagram in figure 3,
where the virtual graviton in the loop is attached to the two external gravitons.
\begin{figure}[h]
\begin{center}
\includegraphics{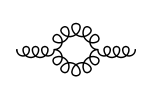}
\end{center}
\caption{\label{fig-tre}Graviton self-energy at one-loop due to a virtual graviton.}
\end{figure}
However, a simple and useful description of the graviton in first quantization is still missing. One may try to use
the $O(4)$ spinning particle of refs. \cite{Gershun:1979fb, Howe:1988ft}, that indeed
describes the propagation of the graviton in flat space. The problem is that it is not known how to couple that model
to an external gravitational background, and the best one could do so far is to couple it to AdS spaces \cite{Kuzenko:1995mg}
and  to conformally flat spaces \cite{Bastianelli:2008nm}.

Given this state of affairs, a less elegant but viable option is that of constructing a  worldline representation of the differential operators
that arise in  the quadratic approximation of the Einstein-Hilbert action. We have recently analyzed the latter approach
in \cite{Bastianelli:2013tsa}, and we review it here.
Then, we use it to compute the one-loop graviton self-energy due to the graviton itself with worldline methods.

\section{The worldline formalism in gravitational backgrounds}\label{sectionWF}

The case of a virtual scalar particle contributing to the graviton self-energy is the simplest one, but  contains already
all the essential elements  that enter in a worldline description of the process. It was treated in \cite{Bastianelli:2002fv}
to exemplify the worldline formalism in a gravitational background.
To review that, one may start recalling that the action of a relativistic scalar particle 
in a spacetime of $D$ dimensions is proportional to the length of its worldline
\be
S[x]=-m \int ds
\label{a1}
\ee
where $ds =d\tau\sqrt{-\dot x^\mu \dot x_\mu} $ with $x^\mu(\tau)$ a generic worldline parametrized by  $\tau$
and $\dot x^\mu =\frac{d x^\mu}{d\tau}$.
The square root appearing in the action can be avoided if one uses an einbein $e(\tau)$ for the intrinsic geometry of the worldline.
The action takes the form  \cite{Brink:1976sz}
\be
S[x, e] =  \int d\tau\, \frac12 (e^{-1} \dot x^\mu \dot x_\mu - e m^2)
\label{a2}
\ee
which has the advantage over (\ref{a1}) of having a smooth massless limit. This action is otherwise
essentially equivalent to  (\ref{a1}), as the equation of motion of the einbein, $ e^{-2} \dot x^2 + m^2=0$, can be solved for  $m\neq 0$
 by taking a positive square root to give  $e= \frac{\sqrt{-\dot x^2}}{m}$. The latter plugged back into   (\ref{a2})
 produces the original action in (\ref{a1}). This model has a reparametrization invariance, that must be gauge-fixed upon quantization.
 One can show that: {\em i)}
quantizing the model  in flat space  and choosing the topology of the worldline to be that of a segment gives the propagator
of the free Klein-Gordon field,
{\em ii)} quantizing the model on a closed worldline gives instead its one-loop effective action.
The complication of the gauge fixing procedure, and the related emergence of an integration over a modular parameter, can be short cut
as the answer must be the same as that obtained long ago by Schwinger  with his proper-time representation of the propagator and one-loop
effective action \cite{Schwinger:1951nm}.

Let us consider the one-loop effective action in a curved background with metric $g_{\mu\nu}$.
We proceed with euclidean conventions obtained after a Wick rotation to euclidean times.
The classical particle action (\ref{a2}), Wick rotated and coupled to the background metric $g_{\mu\nu}$, is
\bea
S[x,e] = \int_{0}^{1} d\tau\, \frac12 \left ( e^{-1} g_{\mu\nu}(x) \dot x^\mu \dot x^\nu + e (m^2 + \xi R(x) ) \right )
\label{a3}
\eea
where an arbitrary non-minimal coupling to the scalar curvature $R$, with coupling constant $\xi$,
has been naturally included (it has the same dimension of the mass term).
Taking the worldline to be a closed loop, i.e. having the topology of the circle $S^1$, and performing the path integral quantization
with the related gauge-fixing procedure,
 one obtains the one-loop effective action  $\Gamma[g]$  in terms of a standard Feynman path integral
with an additional  integration over the Fock-Schwinger proper time $T$
\bea
\Gamma[g]  = -{1\over 2} \int_0^\infty {dT\over T } \int_{S^1} {\cal D}x\ e^{-S_{gf}[x]}
\label{ea}
\eea
where
\bea
S_{gf}[x] = \int_{0}^{1} d\tau \left ( {1\over 4 T} g_{\mu\nu}(x) \dot x^\mu \dot x^\nu +  T  (m^2 + \xi R(x) )\right )
\label{gfa}
\eea
is the action (\ref{a3}) evaluated in the gauge $e(\tau)=2 T$. The path integration is over functions on the circle, i.e. periodic functions.
Up to the normalization appropriate for a real (uncharged) scalar particle,
and up to the integration over the proper time $T$, one finds a path integral
identical to that of a non-relativistic particle (of mass $M=\frac{1}{2T}$)
in a curved space (i.e. a one-dimensional nonlinear sigma model) with additional scalar potential interactions.

Thus, one is left with a path integration over the coordinates $x^\mu $.
This is non-trivial as the path integral for the nonlinear sigma model in (\ref{gfa})
needs a regularization for making it well-defined.  The regularization must be introduced to
fix certain ambiguities that arise in the perturbative calculation of the path integral.
These ambiguities take the form of products of distributions, which are ill defined in the absence of a regularizing procedure.
They are akin to the ordering ambiguities of the canonical quantization of the model.
Such path integrals for nonlinear sigma models have been used to evaluate trace
anomalies for quantum field theories in 2, 4 and 6 dimensions
\cite{Bastianelli:1992be,Bastianelli:1993ct,Bastianelli:2000dw},
and in that context three different regularizations have been analyzed and applied:
mode regularization  (MR) \cite{Bastianelli:1992be, Bastianelli:1998jm, Bastianelli:1998jb}
time slicing (TS)  \cite{DeBoer:1995hv, deBoer:1995cb},
and dimensional regularization (DR) \cite{Bastianelli:2000nm}.
The DR regularization with the correct covariant counterterm was developed after the results
of \cite{Kleinert:1999aq} and \cite{Bastianelli:2000pt},
which dealt with nonlinear sigma models in the infinite propagation-time limit.
All these regularizations require different counterterms to produce
the same physical results. The counterterms needed for nonlinear sigma models with $N$ extended supersymmetry 
have been  worked out in \cite{Bastianelli:2011cc} for all the regularization schemes mentioned above. They are needed for treating 
particles of spin $\frac{N}{2}$, and used for that purpose in \cite{Bastianelli:2012bn}. For the present applications only the cases $N=0,1,2$ are relevant.

A final comment on the path integral measure.
The  covariant measure in (\ref{ea}) takes the form
\be
{\cal D} x = Dx \prod_{ 0\leq \tau < 1} \sqrt{\det g_{\mu\nu}(x(\tau))}  \;,  \qquad Dx=\prod_{ 0\leq \tau < 1}  d^Dx(\tau)
\ee
and can be represented more conveniently by introducing
bosonic $a^\mu$ and fermionic $b^\mu , c^\mu$ ghosts
\bea
{\cal D} x =Dx \int { D} a { D} b { D} c \; e^{- S_{gh}[x,a,b,c]}
\eea
where
\bea
S_{gh}[x,a,b,c]
= \int_{0}^{1} d\tau \; {1\over 4T}g_{\mu\nu}(x)(a^\mu a^\nu
+ b^\mu c^\nu)
\label{ghac}
\eea
so that all measures are translational invariant, a property useful for developing the perturbative expansion around a free gaussian theory.
Upon regularization one may note that vertices arising from the ghost action (\ref{ghac}) help to cancel potential infinities, so that the
counterterms of the various regularization schemes are finite.
Details on the construction and applications of these path integrals in curved space may be found in the book \cite{Bastianelli:2006rx}.

The path integral in (\ref{ea}) is on periodic functions (functions on $S^1$), so that it naturally computes a trace in the Hilbert  space
of the particle
\be
\int_{S^1} {\cal D}x\ e^{-S_{gf}[x]} = {\rm Tr}\; e^{-T \hat H}
\ee
where the quantum hamiltonian $\hat H$ arises from the canonical quantization of $S_{gf}$, and is represented  by
the Klein-Gordon operator $(-\Box +m^2 +\xi R)$. This way one recognizes  (\ref{ea})
as the Schwinger formula for the euclidean one-loop effective action $\Gamma[g]$
\be
\Gamma[g] = {1\over 2} {\rm Tr} \log  (-\Box +m^2 +\xi R) =
-{1\over 2} \int_0^\infty {dT\over T }\; {\rm Tr}\; e^{-T (-\Box+m^2 +\xi R)}
\ee
obtained by quantizing the action of a real Klein-Gordon field $\phi$  coupled to gravity
\bea
S[\phi,g]
= \int d^Dx \sqrt{g}\, {1\over 2}
(g^{\mu\nu} \partial_\mu\phi \partial_\nu\phi +m^2\phi^2 +\xi R\phi^2 )
\label{uno}
\eea
through a field theoretical path integral ($ e^{-\Gamma[g]} = \int {\cal D}\phi\ e^{-S[\phi,g]}= {\rm Det}^{-\frac12}  (-\Box+m^2 +\xi R)$).

\section{The graviton self-energy due to particles of spin $s\leq 1$}\label{sectionSELow}

We are ready to explicitate the calculation of the self-energy diagrams by expanding the worldline representation
of the effective action to second order in the metric fluctuations $h_{\mu\nu}= g_{\mu\nu}-\delta_{\mu\nu}$.
For definiteness, we adopt the DR scheme which  requires the covariant counterterm
\be
\Delta S_{DR}[x] = - \int_{0}^{1} d\tau\, \frac{T}{4}R(x) \;.
\ee
Collecting all terms, the formula for the gravitational effective action induced by a spin 0 particle
in the worldline representation is given by
\bea
\Gamma[g]
= - {1\over 2} \int_0^\infty {dT\over T } \int Dx Da Db Dc
\ e^{-S}
\label{ea2}
\eea
with
\bea
S= \int_{0}^{1} d\tau  \left ( {1\over 4 T}  g_{\mu\nu} (\dot x^\mu \dot x^\nu + a^\mu a^\nu +b^\mu c^\nu)
+  T (m^2 + \bar \xi  R) \right )
\label{gfa2}
\eea
where
$\bar \xi =\xi -{1\over 4} $ includes the DR counterterm. This path integral representation may be expanded to second
order in the metric fluctuation $h_{\mu\nu}= g_{\mu\nu}-\delta_{\mu\nu}$ to obtain
the contribution to the graviton self-energy.
One obtains directly its Fourier transform in momentum space by substituting  in (\ref{gfa2})
$g_{\mu\nu}=\delta_{\mu\nu}+h_{\mu\nu}$  with the fluctuations given by the sum of two plane waves
\be
h_{\mu\nu} (x) = \sum_{i=1}^{2} \epsilon_{\mu\nu}^{(i)}  e^{ik_i \cdot x}
\label{plane}
\ee
and picking up the terms linear in each polarization $\epsilon^{(i)}_{\mu\nu} $. We denote the resulting contribution to the self-energy by
 $(2\pi)^D \delta^D(k_1+k_2)\Gamma_{(k_1,k_2)}$, anticipating momentum conservation.

The detailed calculation may be found in  \cite{Bastianelli:2002fv}, though the salient points are the following ones.
One considers first the case with $\bar \xi=0$.
From expanding the metric in the kinetic term one finds vertex operators for the emission (or absorption) of one graviton
of the form
\be
V[\epsilon, k]=\int_{0}^{1} d\tau  {1\over 4 T}  \epsilon_{\mu\nu} (\dot x^\mu \dot x^\nu + a^\mu a^\nu +b^\mu c^\nu) e^{ik \cdot x}
\ee
and the self-energy is obtained by calculating the correlation
functions of two such operators in the free gaussian theory with $g_{\mu\nu}=\delta_{\mu\nu}$.
The zero mode (i.e. the constant part)  of the quantum variables $x^\mu(\tau)$ can be separated from the path integral, and its integration
produces immediately momentum conservation. Subtleties for factoring out the zero mode in curved space have been discussed extensively in
\cite{Bastianelli:2003bg}, however those issues are not crucial and can be avoided here as the calculation at the end 
is performed in the flat space limit. Thus, the remaining path integration can be carried out using Wick contractions to obtain
\bea
\Gamma_{(k,-k)}
= - {1\over 8}  {1\over (4 \pi)^{D\over 2}}
\int_{0}^{\infty}
{dT\over T^{1 + {D\over 2}}} e^{-m^2 T} \left (R_1 I_1 + R_2 I_2 - 2 T k^2 (R_3 I_3 - R_4 I_4) + 4 T^2 k^4 R_5 I_5 \right )
\eea
where $k=k_1=-k_2$ and
$R_i = \epsilon^{(1)}_{\mu\nu}
R_i^{\mu\nu\alpha\beta} \epsilon^{(2)}_{\alpha \beta} $
with
\bea
R_1^{\mu\nu\alpha\beta} \!\! &=& \!
\delta^{\mu\nu}\delta^{\alpha\beta}  \;, \qquad \qquad \qquad \qquad \quad \ \ \;
R_2^{\mu\nu\alpha\beta} = \delta^{\mu\alpha}\delta^{\nu\beta}+ \delta^{\mu\beta}\delta^{\nu\alpha}
\nonumber \\
R_3^{\mu\nu\alpha\beta} \!\! &=& \!
{1\over k^2} \, (\delta^{\mu\alpha} k^\nu k^\beta +
\delta^{\nu\alpha} k^\mu k^\beta +
\delta^{\mu\beta} k^\nu k^\alpha +
\delta^{\nu\beta} k^\mu k^\alpha )
\nonumber \\
R_4^{\mu\nu\alpha\beta} \!\! &=& \!
 {1\over k^2} \, (
\delta^{\mu\nu} k^\alpha k^\beta
+\delta^{\alpha\beta} k^\mu k^\nu) \;, \qquad \quad
R_5^{\mu\nu\alpha\beta} =
 {1\over k^4}\, k^\mu k^\nu k^\alpha k^\beta
\eea
while the integrals  in the correlation functions are calculated in  DR to the values
\bea
I_1 \!\! &=& \!\!
\int_0^1  d\tau \ e^{- T k^2 (\tau-\tau^2)} \nonumber \;, \qquad  I_2 =  {1\over 4} T k^2  -2 + I_1 \;, \qquad
I_3 =  {1\over 8} -{1\over 2 T k^2 }(1-I_1) \nonumber \\[1mm]
I_4 \!\! &=& \!\! {1\over 2 T k^2 }(1-I_1) \;, \qquad I_5 =  {1\over 8 T k^2 }-{3\over 4 T^2 k^4}(1-I_1) \ .
\eea
The calculation is valid for arbitrary spacetime dimension $D$, which may be extended to complex values
in view of spacetime renormalization.  Carrying out the proper time integral
one finds
\bea
(4 \pi )^{D\over 2} \Gamma_{(p,-p)} \!\! &=& \!\! -{1\over 8} \Gamma(-\tfrac D2)
\left  ( (K^2)^{{D\over 2}} (R_1 + R_2 - R_3 - R_4 + 3 R_5 )
- (m^2)^{{D\over 2}} (2 R_2 - R_3 - R_4 + 3 R_5 ) \right )
\nonumber \\[2mm]
\!\! &-& \!\!
{1\over 32}
\Gamma(1-\tfrac D2)  k^2 (m^2)^{{D\over 2}-1} ( R_2  - R_3 + 2 R_5 )
\label{spin0}
\eea
where we have used the definition
\be
(K^2)^x = \int_0^1 d\tau\, (m^2 + k^2 (\tau -\tau^2))^x \ .
\ee

Additional terms $\Delta \Gamma_{(k,-k)} $ are
present in the case with $\bar\xi\neq 0$. First, there are corrections to the vertex operator for the emission of a single graviton
arising from expanding to the linear order the scalar curvature in (\ref{gfa2}).
Secondly, there is a vertex operator for the emission
of two gravitons obtained  by expanding the scalar curvature to second order. Their inclusion produces
the following additional contribution to (\ref{spin0})
\bea
 (4 \pi )^{D\over 2} \Delta \Gamma_{(k,-k)} \!\! &=& \!\!
-{\bar\xi  \over 8}  \Gamma(1-\tfrac D2) k^2   \left ( (m^2)^{{D\over 2}-1}
 (2 R_1  +  R_2 -  R_3 - 2 R_4  + 4  R_5 ) \right .
\nonumber \\[2mm]
&& \hspace{-2cm} \left .
-4 (K^2)^{{D\over 2}-1} ( R_1  -  R_4  +  R_5 ) \right )
-{\bar\xi^{\,2} \over 2}  \Gamma(2-\tfrac D2) k^4 (K^2)^{{D\over 2}-2} (R_1 - R_4  +R_5) \ .
\eea
This gives the final regulated (but not renormalized) self-energy contribution from spin 0.

The final two-point function can be written in a more compact form using the tensors
\be
S_1 = R_1 -  R_4 + R_5 \;, \qquad
S_2 = R_2 -  R_3 + 2 R_5
\ee
which satisfy
$k_\mu S_1^{\mu\nu\alpha\beta} = k_\mu S_2^{\mu\nu\alpha\beta} = 0$ (when the polarizations are stripped off), 
and make it easier to check the gravitational Ward identities.
For arbitrary $\bar \xi$ it reads
\bea
&& \Gamma^{\mbox{\tiny (0)}}_{(k,-k)} =
-{\Gamma(-\tfrac D2) \over 8 (4 \pi )^{D\over 2} }  \left ( (m^2)^{{D\over 2}} ( R_1  -R_2 )
+ ((K^2)^{{D\over 2}} -(m^2)^{{D\over 2}})(S_1  +S_2)
\right ) \nonumber \\[2mm] && \qquad
- {\Gamma(1-\tfrac D2) \over 32 (4 \pi )^{D\over 2} }  
\left [ k^2  (m^2)^{{D\over 2}-1} S_2 +4\bar\xi  k^2 \left ( (m^2)^{{D\over 2}-1} ( 2 S_1  + S_2 ) - 4 (K^2)^{{D\over 2}-1}  S_1 
\right )\right ]
\nonumber \\[2mm] && \qquad
-{\Gamma(2-\tfrac D2) \over 2 (4 \pi )^{D\over 2} }  \bar\xi^{\,2} 
k^4
(K^2)^{{D\over 2}-2} S_1  \ .
\label{se-0}
\eea
The value $\bar \xi= -{1\over 4}$ ($\xi=0$)
describes the self-energy from a scalar with minimal coupling.
A conformally coupled scalar needs instead the
value $\bar \xi = {1\over 4(1-D)}$ (i.e. $\xi = {(D-2)\over 4(D-1)}$)
together with $m^2 =0$.
Finally, the value
$\bar \xi= 0$ ($\xi={1\over 4}$)
allows for  the simplest computation in the worldline formalism
as vertex operators may contain one graviton only.

For a Dirac particle of spin 1/2 one proceeds in a similar way, using the spinning particle of \cite{Brink:1976sz}.
 From the worldline point of view this amounts to supersymmetrize the previous result for $\bar \xi= 0$, and one finds \cite{Bastianelli:2002qw}
\bea
\Gamma^{\mbox{\tiny ($\tfrac12$)}}_{(k,-k)} \!\! & = &\!\!
{ 2^{{D \over 2}}\over 8 (4 \pi )^{D\over 2} } \left [ \Gamma(-\tfrac D2)
  \left ( (m^2)^{{D\over 2}} ( R_1  -R_2 -S_1 -S_2)
  + (K^2)^{{D\over 2}}(S_1  +S_2) \right )  \right.
                \nonumber \\
 &&\qquad \quad\ +  \left.   \frac14 \Gamma(1-\tfrac D2)
 k^2 (K^2)^{{D\over 2}-1} S_2  \right ] \ .
 \label{se-12}
\eea

Similarly, for a spin 1 particle one may use the N=2 extended spinning particle model. It  has been used in
\cite{Bastianelli:2005uy} (see also \cite{Bastianelli:2005vk})
to find the contribution from massless and massive $p$-forms in arbitrary dimensions. In $D=4$ the only new result
with respect to the previous ones (up to dualities)  is that of a particle of spin 1, whose contribution reads
\bea
\Gamma^{\mbox{\tiny (1)}}_{(k,-k)} &=& N_{dof}  \Gamma^{\mbox{\tiny (0)}}_{(k,-k)}
-{1 \over 8 (4\pi)^{D \over 2}} (S_2-2S_1)  
\nonumber \\ && \times 
\left [ \Gamma(1-\tfrac D2) k^2  \Big{(}2 (K^2)^{{D \over 2} -1} - (m^2)^{{D \over 2} -1}\Big{)}
+ \frac12  \Gamma(2-\tfrac D2) k^4
(K^2)^{{D \over 2}-2} \right ] \quad\quad
\label{se-1}
\eea
where $\Gamma^{\mbox{\tiny {0}}}_{(k,-k)}$ is the two-point function due to a minimally coupled scalar 
($\bar \xi= -{1\over 4}$, i.e. $\xi=0$), while  $N_{dof}=2$ for a massless spin 1 particle 
and $N_{dof}=3$ for a massive one.

These  worldline results agrees with those computed with standard Feynman rules, 
though it may be noticed how the worldline computation produces simpler
and more compact expressions. Early calculations of the graviton self-energy 
may be found in \cite{Capper:1973bk}, \cite{Capper:1973mv}, \cite{Capper:1974ed},
where one finds the contributions   due to a  massive scalar with minimal coupling, a massless fermion, and the photon, respectively.
In \cite{Capper:1973pv} one finds a calculation of the graviton self-energy due to the graviton itself, 
which is the most tricky one to obtain with worldline methods.

\section{Graviton self-energy due to spin 2 loop}

In this section we are going to review the model developed in \cite{Bastianelli:2013tsa} to produce one-loop quantum gravity amplitudes in four dimensions, and present the computation of the graviton self-energy induced by a graviton loop. As mentioned in the introduction, 
a worldline description of gravitons is not easy to achieve. The $O(4)$ extended spinning particle indeed describes the free propagation 
of a spin two particle in conformally flat spacetimes, but its coupling to general curved backgrounds is prevented by obstructions in the 
constraint algebra defining the model \cite{Bastianelli:2008nm}.
A viable alternative, though less elegant, to reproduce one-loop graviton contributions has been developed in \cite{Bastianelli:2013tsa}: its starting point is the quadratic expansion, in background field method, of the Einstein-Hilbert action. After fixing the quantum gauge symmetry in Fock-De Donder gauge, the one-loop effective action is given in terms of functional traces as
\begin{equation}\label{gravity EA}
\Gamma[g]=\Gamma_{\scalebox{0.5}{TT}}+\Gamma_{\scalebox{0.5}{S}}-2\Gamma_{\scalebox{0.5}{V}}\;,
\end{equation}
where the three pieces represent contributions from traceless tensor fluctuations, scalar trace fluctuations and vector ghosts, respectively. Each contribution to the effective action is given by a functional trace of the form $\Gamma={\rm Tr}\ln[\Pi\,\Box+{\cal R}]$, where $\Pi$ is a tensor projector for the given representation and ${\cal R}$ stands for curvature couplings. By means of the Schwinger exponentiation of logarithms with the proper time, each functional trace in \eqref{gravity EA} can be represented as a quantum mechanical partition function, as reviewed for the scalar in section \ref{sectionWF}. 
At this stage this is the well-known heat kernel method developed by DeWitt for obtaining the one-loop effective action for quantum gravity \cite{DeWitt:1965jb}.
The goal of \cite{Bastianelli:2013tsa} was  to engineer suitable worldline actions able to reproduce the various contributions 
of \eqref{gravity EA} as their partition functions, namely
\begin{equation}\label{EA Wline}
\begin{split}
\Gamma_{\scalebox{0.5}{TT}} &= -\frac12\int_0^\infty\frac{dT}{T}e^{-m^2T}\int_0^{2\pi}\frac{d\phi}{2\pi}e^{\frac72 i\phi}\int_P\mathcal{D}x\int_AD\bar\psi D\psi\, e^{-S_{\scalebox{0.5}{TT}}[x,\bar\psi,\psi;\phi]}\;,\\[2mm]
\Gamma_{\scalebox{0.5}{V}} &= -\frac12\int_0^\infty\frac{dT}{T}e^{-m^2T}\int_0^{2\pi}\frac{d\phi}{2\pi}e^{i\phi}\int_P\mathcal{D}x\int_A D\bar\lambda D\lambda \,e^{-S_{\scalebox{0.5}{V}}[x,\bar\lambda,\lambda;\phi]}\;,\\[2mm]
\Gamma_{\scalebox{0.5}{S}} &= -\frac12\int_0^\infty\frac{dT}{T}e^{-m^2T}\int_P\mathcal{D}x\,e^{-S_{\scalebox{0.5}{S}}[x]}\;,
\end{split}
\end{equation}
where the modular integrals consist of integration over the proper time $T$ and $U(1)$ modulus $\phi$, with the latter ensuring projection onto the desired sectors of the particle Hilbert 
spaces\footnote{This projection mechanism was observed in \cite{Bastianelli:2005vk}
and applied often in worldline applications, as in \cite{Bastianelli:2013pta} for projecting to some irreducible color degrees of freedom, and
 in \cite{Bastianelli:2011pe,Bastianelli:2012nh}  to achieve the description of quantum $(p,q)$-forms on K\"ahler spaces.},
 as explained in detail in \cite{Bastianelli:2013tsa}. The subscripts $A$ and $P$ stand for (anti)-periodic boundary conditions, and the fictitious mass $m$ is an infrared regulator that will be eventually set to zero.
The worldline actions appearing in \eqref{EA Wline} read
\begin{equation}\label{wline actions}
\begin{split}
S_{\scalebox{0.5}{TT}}[x,\bar\psi,\psi;\phi] &= \int_0^1d\tau\,\Big[\,\frac{1}{4T}\,g_{\mu\nu}\,\dot x^\mu\dot x^\nu+\frac12\,\bar\psi_{ab}\left(\partial_\tau+i\phi\right)\psi^{ab}+\omega_{\mu ab}\,\dot x^\mu\bar\psi^a\cdot\psi^b\\
&\qquad\qquad-T\left(R_{abcd}\,\psi^{ac}\bar\psi^{bd}+R_{ab}\,\psi^a\cdot\bar\psi^b+\tfrac34\,R\right)\Big]\;,\\[2mm]
S_{\scalebox{0.5}{V}}[x,\bar\lambda,\lambda;\phi] &= \int_0^1d\tau\,\Big[\,\frac{1}{4T}\,g_{\mu\nu}\,\dot x^\mu\dot x^\nu+\bar\lambda_a\left(\partial_\tau+i\phi\right)\lambda^a+\omega_{\mu ab}\,\dot x^\mu\lambda^a\bar\lambda^b-T\left(R_{ab}\,\lambda^a\bar\lambda^b+\tfrac34\,R\right)\Big]\;,\\[2mm]
S_{\scalebox{0.5}{S}}[x] &= \int_0^1d\tau\,\Big[\,\frac{1}{4T}\,g_{\mu\nu}\,\dot x^\mu\dot x^\nu-\frac{T}{4}\,R\Big]\;,
\end{split}
\end{equation}
where the worldline fermions $\psi^{ab}(\tau)$, $\bar\psi^{ab}(\tau)$ are symmetric traceless tensors in spacetime: $\psi^{ab}=\psi^{ba}$, $\psi^a_a=0$, with $a,b,..$ being four dimensional flat Lorentz indices, while the fermions $\lambda^a(\tau)$ and $\bar\lambda^a(\tau)$ are spacetime vectors. The above actions, being nonlinear sigma models, require regularization, and the corresponding DR counterterms found in \cite{Bastianelli:2013tsa} are already included in \eqref{wline actions}. 
At this stage one can expand the background metric\footnote{Since the vielbein and spin connection are present, one actually expands those in plane waves and, after the calculation is performed, one goes back to the metric basis.} in plane waves around flat space in order to compute contributions to the $n$-point functions in momentum space. For the self-energy one sets $g_{\mu\nu}(x)=\delta_{\mu\nu}+\sum_{i=1}^2\epsilon_{\mu\nu}^{(i)}e^{ik_i\cdot x}$ and keeps the terms linear in $\epsilon^{(1)}\epsilon^{(2)}$. The calculation is very much akin to the one performed for the spin one loop, due to the extra angular integrals with respect to spin one half and spin zero cases, and we shall present only the final result, that is given by summing the three contributions as dictated by \eqref{gravity EA}
\begin{equation}\label{graviton metric}
\begin{split}
\Gamma^{\mbox{\tiny (2)}}(k,-k)&=-\frac{1}{4(4\pi)^{D/2}}\Big [\Gamma(-\tfrac D2)\Big(\left(K^2\right)^{\frac D2}(S_1+S_2)+(m^2)^{\frac D2}(R_1-R_2-S_1-S_2)\Big)\\
&+k^2\Gamma(1-\tfrac D2)\Big(\left(K^2\right)^{\frac D2-1}(4S_2-15S_1)+\tfrac{15}{2}(m^2)^{\frac D2-1}S_1\Big)\\
&+k^4\Gamma(2-\tfrac D2)\Big(\left(K^2\right)^{\frac D2-2}(S_2-\tfrac74\,S_1)\Big)\Big]\;,
\end{split}
\end{equation}
where we have used the same conventions of the previous section for tensor structures.
It is possible now to remove the infrared regulator by sending $m^2\to0$. In this limit the factors of $K^2$ yield Euler beta functions, since
$$
\lim_{m^2\to0}\left(K^2\right)^x=k^{2x}\int_0^1d\tau\,[\tau(1-\tau)]^x=k^{2x}B(x+1,x+1)\;,
$$
and we can write the dimensionally regulated (in spacetime) result as
\begin{equation}\label{graviton massless}
\begin{split}
&\Gamma^{\mbox{\tiny (2)}}(k,-k)=-\frac14\left(\frac{k^2}{4\pi}\right)^{\frac D2}\Big[
\Gamma(-\tfrac D2)B(\tfrac D2+1,\tfrac D2+1)(S_1+S_2)\\[2mm]
&\qquad +\Gamma(1-\tfrac D2)B(\tfrac D2,\tfrac D2)(4S_2-15S_1)+\Gamma(2-\tfrac D2)B(\tfrac D2-1,\tfrac D2-1)(S_2-\tfrac74\,S_1)\Big]\;.
\end{split}
\end{equation}
In order to display the physical result in four dimensions we take $D=4-2\varepsilon$, and use the mass 
scale $\mu$ of dimensional regularization  as in the $\overline{\rm MS}$ scheme 
(note also that we have kept the  gravitational coupling constant absorbed in the polarisation tensor).
The final result reads
\begin{equation}\label{graviton ms bar}
\begin{split}
&\Gamma^{\mbox{\tiny (2)}}(k,-k)=-\frac{k^4}{64\pi^2}\left[\left(\frac1\varepsilon-\ln\frac{k^2}{\mu^2}\right)
\left(\frac{7}{20}\,S_2+\frac{23}{30}\,S_1\right)+\frac{41}{150}\,S_2+\frac{724}{225}\,S_1\right]\;.
\end{split}
\end{equation}
One can define linearized curvature invariants in momentum space by using the plane waves in \eqref{plane}
\begin{equation}
\begin{split}
R^2_{\mu\nu}(k) &= R^{({\rm lin})}_{\mu\nu}(\epsilon_1 e^{ikx}) R_{({\rm lin})}^{\mu\nu}(\epsilon_2 e^{-ikx}) = \frac18 k^4(2S_1+S_2)\\[2mm]
R^2(k) &= R^{({\rm lin})}(\epsilon_1 e^{ikx}) R^{({\rm lin})}(\epsilon_2 e^{-ikx}) = k^4\,S_1
\end{split}
\label{33}
\end{equation}
so that we can rewrite \eqref{graviton ms bar} in a more suggestive form, that makes manifest the relation with the effective action in configuration space
\begin{equation}\label{grxaviton ms curvatures}
\begin{split}
 \Gamma^{\mbox{\tiny (2)}}(k,-k)&=-\frac{1}{8\pi^2}\left[\left(\frac1\varepsilon-\ln\frac{k^2}{\mu^2}\right)
\left(\frac{7}{20}\,R^2_{\mu\nu}(k)+\frac{1}{120}\,R^2(k)\right) 
+\frac{41}{150}\,R^2_{\mu\nu}(k)+\frac{601}{1800}\,R^2(k)\right ]\;,
\end{split}
\end{equation}
where one can easily recognize the well known logarithmic divergencies of Einstein gravity in four dimensions \cite{'tHooft:1974bx}.

The relevant log term  at large $k^2$ can be extracted also for the other particles in the loop.
From eqs. (\ref{se-0}), (\ref{se-12}) and (\ref{se-1}) one finds the following additional contribution
to the graviton self-energy 
\be
\Gamma(k,-k)=\frac{1}{8\pi^2} \ln\frac{k^2}{\mu^2}
\left [ \frac{R^2_{\mu\nu}}{120} \left (N_{\mbox{\tiny 0}} + 6 N_{\mbox{\tiny $\tfrac12$}} + 12 N_{\mbox{\tiny 1}}\right )
+ \frac{R^2}{240}\Big(  (1-20\xi + 60 \xi^2) N_{\mbox{\tiny 0}} -4 N_{\mbox{\tiny $\tfrac12$}} -8 N_{\mbox{\tiny 1}}\Big )
 \right]
 \label{35}
\ee
 where $N_{\mbox{\tiny 0}}$,  $N_{\mbox{\tiny $\tfrac12$}}$ and  $N_{\mbox{\tiny 1}}$, are the number of 
particles of spin 0, $\frac12$ and 1, respectively. For simplicity, we have taken all scalars with the same non-minimal coupling $\xi$
and considered the spin $\frac12$ particles as Dirac fermions.
These particles are all taken massless, as for large enough  $k^2$ the mass can be disregarded, 
so that a massive spin 1 particle would count as a massless one plus a minimally coupled scalar.\\
As a check on \eqref{35}, one may note that the same expression multiplying  $\ln\frac{k^2}{\mu^2}$ sits also in the diverging part,  
multiplied by $-\frac1\varepsilon$ as in \eqref{grxaviton ms curvatures}.
For conformal fields this expression gives the conformal anomaly. Indeed, setting $\xi=\frac16$ 
as appropriate for a conformal scalar in four dimensions, one finds that the two tensor structures combine to form the square of the Weyl tensor
(one must take into account  that the topological Euler density is a total derivative, and
vanishes in the plane wave basis considered in \eqref{33}, so that a term $R^2_{\mu\nu\lambda\sigma}$
can be expressed  in function of $R^2_{\mu\nu}$ and $R^2$). 
Thus one can read off the correct conformal anomaly coefficient, the one that depends
on the Weyl tensor squared.

\section*{Acknowledgments} We thank Christian Schubert for valuable comments on the manuscript.

\end{document}